\documentclass[preprint,showpacs,showkeys,aps,endfloats*]{revtex4}

\usepackage{graphicx}
\usepackage{dcolumn}
\usepackage{bm}

\begin{document}

\preprint{APS/123-QED}

\title{Selective Optical Charge Generation, Storage and Readout in a Single Self Assembled Quantum Dot}

\author{D. Heiss}
\author{V. Jovanov}
\author{M. Caesar}
\author{M. Bichler}
\author{G. Abstreiter}
\author{J. J. Finley}
\email{finley@wsi.tum.de}
\affiliation{Walter Schottky Institut, Technische Universit\"at M\"unchen, Am Coulombwall 3, D-85748 Garching, Germany}
\date{\today}
\begin{abstract}
We report the investigation of a single quantum dot charge storage device. The device allows selective optical charging of a single dot with electrons, storage of these charges over timescales much longer than microseconds. Reliable readout of the charge occupancy is realized by time gated photoluminescence technique. This device enables us to investigate the tunneling escape of electrons at high electric fields up to several microseconds and, therefore, demonstrates that with more elaborate pulse sequences such structures can be used to investigate charge and spin dynamics in single quantum dots.   
\end{abstract}
\maketitle
In the context of quantum information processing great progress has been made in the control and investigation of the charge and spin states of single quantum dots in the recent years. In particular the use of quantum point contacts has dramatically enhanced the level of sophistication of experiments on electrostatically defined quantum dots. \cite{Hanson07} Such measurements have revealed the orbital shell structure \cite{Elzerman03} and allowed measurements of single electron spin relaxation and coherence times.\cite{Petta05, Koppens06} For self assembled quantum dots such sensitive charge detection techniques are not yet readily available and other techniques have been employed to probe the charge and spin states. Approaches based on spin to photon polarization conversion have demonstrated spin lifetime measurements in ensembles \cite{Kroutvar04, Heiss07}, however, all optical spin readout on single self assembled QDs is extremely challenging. More recently, sensitive techniques based on direct absorption \cite{Hoegele05,Ramsay08} or time resolved Faraday \cite{Atatuere07} and Kerr \cite{Berezovsky06} rotation have been applied to test the spin of an isolated electron. These techniques allow to deduce properties of spin dynamics like relaxation or dephasing in QDs, but do not include a storage phase, where reliable spin manipulation can be performed over timescales bigger than a few nanoseconds.\\    

In this letter we present an optical method to control and probe the charge state of a single self assembled quantum dot. We charge the dot with a \textit{single} electron via quasi-resonant optical excitation in its p-shell, offering a better control over the charge state than the previous experiments performed with excitation in the wetting layer \cite{Heiss08}. Furthermore, compared to reference~\cite{Heiss08}, the readout fidelity is increased by switching to quasi-resonant optical excitation, which allows us to directly measure the tunneling escape time of the electron on a timescale of microseconds. This kind of reliable optical charge detection should make optical investigations of spin dynamics over extremely long timescales extending into the microsecond regime \cite{Heiss08}.\\

In order to perform optical charging experiments on a single self assembled QD precise control of the local electric field is necessary. This can be achieved by embedding the dot in the intrinsic region of a Schottky photodiode structure formed by a heavily n-doped back contact and a 5~nm thick, semi transparent Ti top contact. The intrinsic region has a total thickness of 140~nm with a single layer of self assembled InGaAs QDs positioned 40 nm above the n-doped layer. This sample structure leads to a static electric field of 70~kV/cm, per volt applied in the intrinsic region of the Schottky diode and a flat band condition ($|$F$|$=0~kV/cm) at an applied voltage of V$_{app}$=0.9~V. An opaque Au shadow mask is evaporated onto the sample surface to allow the optical selection of single dots through 1~$\mu$m diameter shadow mask apertures defined using polybeads spin coated onto the sample surface before metallization. By controlling the voltage V$_{app}$ applied to the Schottky gate, the tunneling escape time of electrons and holes from the QD can be continuously tuned. To ensure that the tunneling time for holes ($\tau_h$) is much faster than for electrons ($\tau_e$) and, thus, enable optical charging, an asymmetric Al$_{0.3}$Ga$_{0.7}$As barrier with a width of 20~nm was grown below the QD layer. The schematic band profile of the device investigated when subject to different bias regimes is shown in figure 1a - 1d.\\

\begin{figure}[h]
	\centering
		\includegraphics[width=1\textwidth]{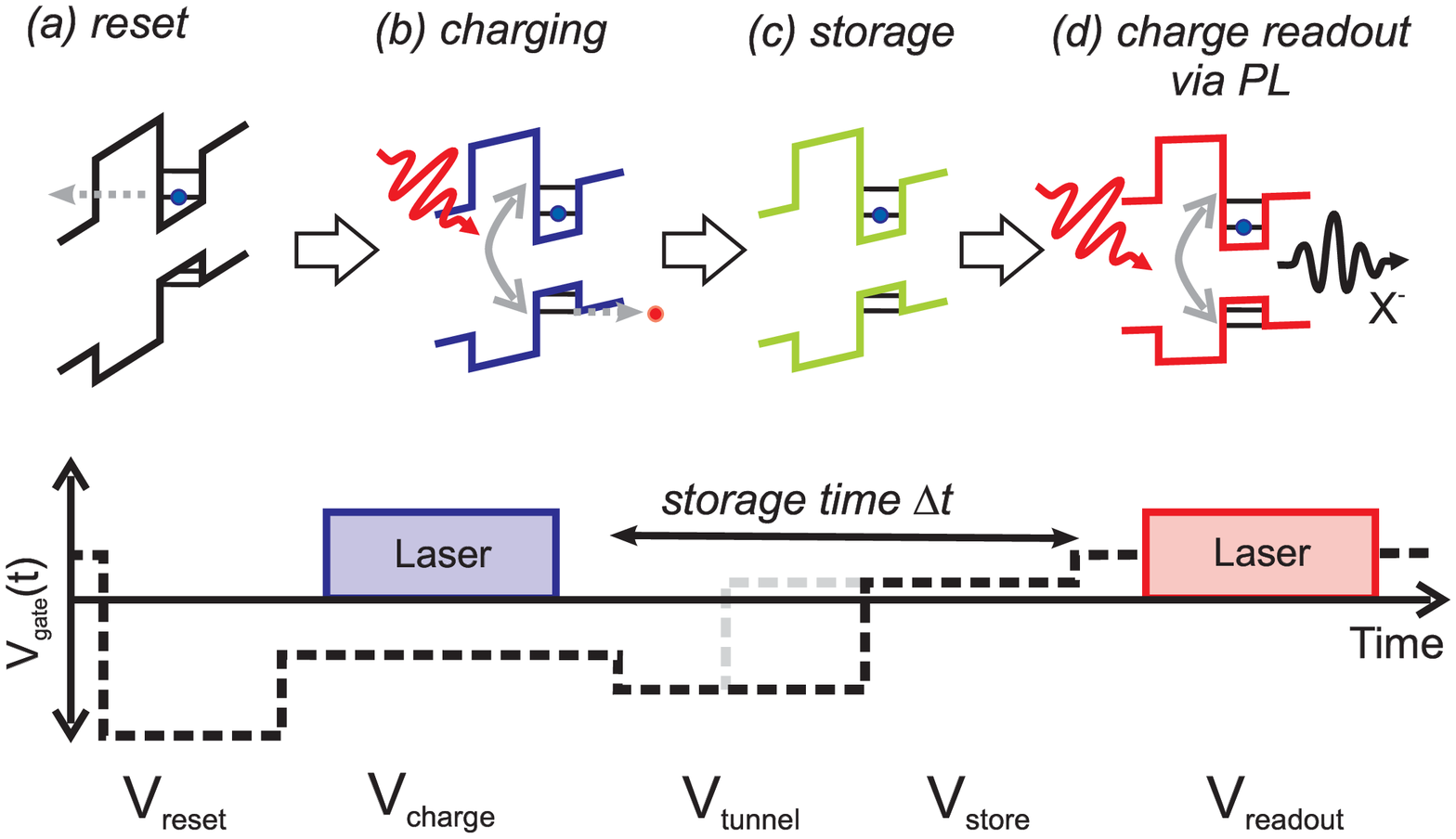}
	\caption{(color online) Schematic representation of the pulse sequence used in the storage experiments. The phases of the experiment are:  (a) reset, (b) selective optical charging, (c) electron storage, (d) charge readout via time gated PL. }
\end{figure}
We begin by introducing the pulse sequence used for our charge storage experiments. A schematic illustration of one measurement cycle is shown in fig 1. It consists of four measurement phases: charge reset (fig 1a), charging (fig~1b), charge storage (fig~1c) and charge readout (fig~1d). Each measurement cycle begins with a 300~ns duration reset pulse (fig. 1a), for which the applied electric field ($|$F$|$=170~kV/cm \cite{Heiss08}) across the QD is chosen such that the electron tunneling time is much faster than 300~ns. This step ensures that the quantum dot is initially uncharged and no cumulative charging effects can take place over several cycles of the experiment. In the next step, shown in fig. 1b, the QD is optically charged. To do this, we illuminate for 300~ns with laser light that is energetically tuned to a discrete p-shell absorption transition. An electric field is applied, such that the hole tunnels towards the Schottky contact faster than the radiative lifetime of the exciton, while the electron stays trapped in the dot due to the presence of the AlGaAs barrier. This charge quickly relaxes to the s-shell. After generation the charge is stored in the dot for a time $\Delta t$ that can be extended over several microseconds. The electric field can be freely modified during that time to test the influence on the stored charge. Finally, the charge state of the dot is tested by time gated photo luminescence (fig~1d). For this, the relative strength of the uncharged (X$^0$) singly (X$^{-1}$) and doubly charged (X$^{-2}$) exciton emission indicates the average charge occupation of n$_e$ = 0, 1 and 2 electrons respectively. It has been shown  that such a charge readout can be performed with high fidelity, for applied electric fields below 7~kV/cm and low optical powers below 10 W/cm$^2$.\cite{Heiss08} \\

The measurements presented in the following were performed using a cryo-microscope in a helium bath cryostat at a temperature of 8~K. The signal was detected using a 0.55~m single spectrometer and a LN$_2$ cooled Si charge coupled device multichannel detector (CCD). We used two separate tunable external cavity Littman-Metcalf diode lasers as excitation sources during the charging and charge detection phases of the measurement. These were gated with acousto-optical modulators to phase lock the illumination and readout pulses to the voltage sequence applied to the Schottky gate. The application of bias voltages to the sample and synchronization with the laser pulses was achieved by a set of arbitrary waveform generators (Agilent AWG33250A).\\
An aperture that showed only emission from a single QD was selected by carefully performing power dependent measurements on a number of apertures and choosing one that exhibited characteristic exciton and multiexciton emission.\\

\begin{figure}[h]
	\centering
		\includegraphics[width=1\textwidth]{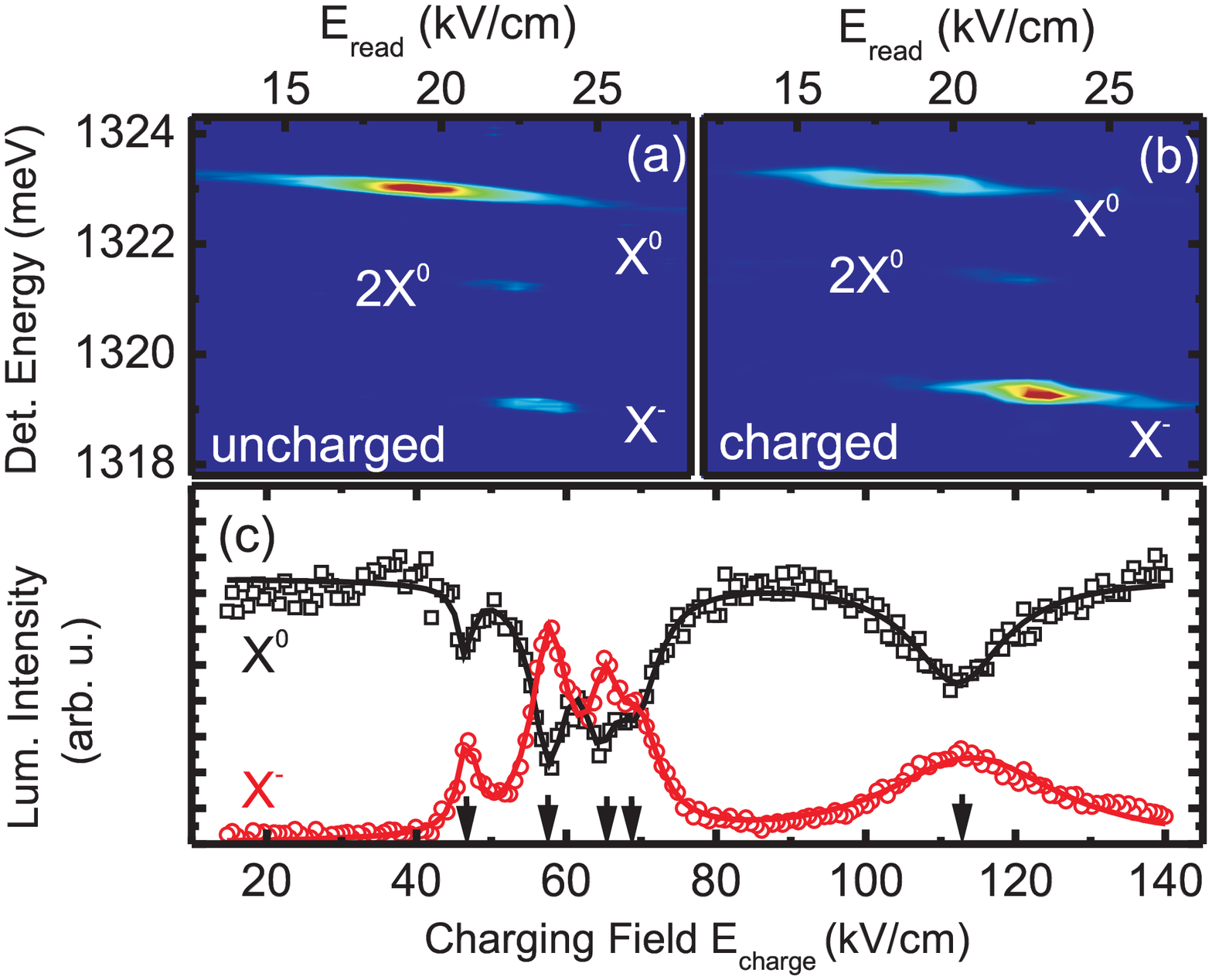}
	\caption{(color online) Photoluminescence - excitation measurements of a (a) charge neutral and (b) a charged QD. (c) Charging of the QD, observed by an increase of the X$^{-}$ intensity, can be achieved through discrete p-shell states of the QD.}
\end{figure}
We continue by discussing the readout phase of our experiment that is of vital importance for high fidelity non-perturbative readout that does not change the charge state of the dot whilst generating luminescence. For this, the rates for optical charging and charge loss due to electron tunneling escape from the dot or single hole capture into the dot must be suppressed. Reliable charge readout up to 100~$\mu$s has been demonstrated in earlier work \cite{Heiss08}, where time gated PL measurements were performed using non-resonant excitation into the wetting layer. For this, the electric field across the dot was set to be close to flat band and low excitation powers of P$_{read}$=1~W/cm$^2$ where used. The fidelity of such measurements can be significantly improved, by pumping into a discrete p-shell absorption line instead of the wetting layer. This arises, since here geminative carrier capture is strongly enhanced, which suppresses charging due to single electron capture and charge loss by recombination with captured single holes during the readout phase of the measurement. Time resolved measurements with p-shell excitation  similar to those presented in \cite{Heiss08} were performed and showed a reduction of the charging and loss rates by approximately a factor of 10 compared to the case of wetting layer excitation. \\

To find a suitable p-shell absorption line we performed PLE-type measurements. The results of these measurements are shown in for a nomainally uncharged dot in fig 2a and for a precharged dot in fig 2b. To avoid power fluctuations, the laser energy was fixed at 1334~meV whilst the p-shell absorption lines are voltage tuned using the quantum confined stark effect (QCSE). From high power PL-measurements recorded as a function of electric field we observed an approximatly  linear stark shift of the p-shell absorption resonances of \~ 5~meV/V. The PLE spectra presented in fig 2a reveal three distinct emission lines. Based on additional electric field and power depended measurements (not shown) we attribute these lines to the neutral exciton (X$^0$), neutral biexciton (2X$^0$) and negatively charged exciton (X$^-$) respectively. In the following considerations we will neglect the 2X$^0$ emission, since it is considerably weaker than the other luminescence under the excitation conditions used in our experiments. Notably, as shown in fig~2, the maximum emission intensity observed for X$^-$ is shifted towards higher electric fields F$_{read}$=23~kV/cm. From the QCSE of the p-shell absorption lines, we can deduce a red shift of the absorption of a negatively charged QD of 0.36~meV in comparrison with the absorption characteristic of a charge neutral dot. We attribute this to a energy-renormalization of a different p-shell absorption line due to the presence of the extra electron charge. This energy shift makes it necessary to divide the readout step into two parts. The readout voltage V$_{read}$ is first set to tune the absorption line into resonance at F$_{read}$=23~kV/cm, which leads to X$^-$ emission if the dot is charged with a single electron, followed by equal amount of time tuned at F$_{read}$=18~kV/cm leading to X$^0$ emission, only if the dot is uncharged. \\ 

In the previous section it has been shown that reliable charge readout is possible and we will continue by discussing the optical charging (fig 1b) of the QD. Two conditions have to be fulfilled for efficient optical charging in the p-shell: firstly the applied electric field F$_{charge}$ is chosen such that tunneling escape of the hole is much faster than the radiative lifetime of the exciton, while the electron escape is much slower than the period of the measurement cycle. For our sample structure this requirement is readily fulfilled in the electric field regime 120 $<$ F$_{charge} <$ 25 kV/cm. Secondly, the energy of the charging laser has to match the energy of the QD discrete p-shell absorption line. To find such absorption resonances the pulse sequence sketched in fig~1 is used, although the storage step (fig~1c)is omitted. The readout phase is depicted in fig~1d and is performed as described in the previous section. During the charging phase (fig~1b) a second laser is gated ON and its energy is kept constant at E$_{Laser}$=1343~meV. To shift the p-shell absorption lines through the laser energy again the QCSE is used. Such a measurement is shown in fig 2c. The X$^0$ and X$^-$ luminescence intensities are plotted as a function of the electric field applied across the QD during the \textit{charging} phase of the experiment. For electric fields above 120~kV/cm and below 25~kV/cm we observe mainly X$^0$ luminescence, since the electric field does not match the charging condition mentioned above. In the mid field regime (120 $<$ F$_{charge} <$ 25 kV/cm) we observe an increase of the X$^-$ luminescence at distinct electric field values accompanied by an anitcorrelated decrease of the X$^0$ intensity. The only parameter varied here is F$_{charge}$ during the charging phase of the experiment, where no luminescence is generated. Therefore, we attribute the anticorrelated resonances in the X$^{0}$/X$^{-}$ intensities observed in the readout phase of the measurement to a change in the electron occupancy of the dot during the charging phase due to optical charging. As shown in fig~2c by full lines the data can be fitted with a set of 5 Lorentzian lines and each resonance is marked with an arrow on the electric field axis. Providing that absorption does not take place in the p-shell of the QD, no charging occurs and we observe mainly X$^0$ emission. As soon as the p-shell absorption lines are tuned by the electric field into resonance with the charging laser, an electron hole pair is created in the p-shell of the dot. The hole will tunnel out of the dot due to the electric field, while the electron is trapped by the AlGaAs barrier and relaxes to the s-shell. Since the electron is stable in the dot, negatively charged trions will be formed during the readout phase of the measurement and the X$^-$ luminescence becomes stronger. A comparrisson of a PLE spectrum is shown in fig 2 off (fig 2a) and on (fig 2b) such a charging resonance. The lorentzian line shape of the resonances in fig 2c resembles the shape of the p-shell absorption lines. Additionally, we observe a lifetime-broadening of the resonances for higher electric fields, since the tunneling time of the hole becomes faster at high fields \cite{Findeis01}. From the line width we can estimate the tunneling time of the hole, which ranges from 5~ps at 47~kV/cm to 0.5~ps at 113~kV/cm.\\

The results demonstrate selective charge initialization and non perturbative readout of electrons in the QD. This method can be the basis for more elaborate investigation of the properties of the selectively generated single electrons. To demonstrate the feasibility of such measurements we directly measured the tunneling escape of a single electron from the QD. We use the storage pulse sequence shown in fig~1, making sure that the QD is charged before storage and that the readout does not alter the charge state of the dot. During the storage phase we apply a low electric field of 25~kV/cm where tunneling escape of electrons does not occur within our storage time. To investigate the time evolution of the escape from the QD, a strong electric field F$_{tunnel} \geq$133~kV/cm is introduced for a time $\Delta t$, where $\Delta t$ can be increased up to 4~$\mu$s. In fig.~3 the fraction of the total luminescense intensity arising from X$^-$ is plotted as a function of $\Delta t$ for the electric fields F$_{tunnel}$=133, 140, 147, and 154~kV/cm. The curves show a clear monoexponential decay from the initial value of 50\% towards 5\%, as the time $\Delta t$ for which F$_{tunnel}$ is applied is increased. Since the X$^-$ intensity directly reflects the probability that the QD is charged with an electron, the decay presented in fig~3 reflects the electric field induced tunneling escape of the electron through the AlGaAs barrier. The extracted tunneling times support this interpretation, since the values are comparable with expected values and the behaviour as a function of electric field is exponential, as expected for tunneling escape from QDs \cite{Fry00}. If we extrapolate the tunneling time to electric fields practicable for storage experiments (F$_{store}$= 30~kV/cm) the electron lifetime becomes very long extending into the range of a few seconds. \\

In summary, we have investigated the charging, charge readout and charge escape on a single selfa ssembled QD. We were able to selectively charge a QD by its p-shell absorption, and reliably read the charge state. Furthermore, we have shown time-resolved optical measurements of the charge state of a single self assembeld QD extending for several microseconds. In the future more elaborate pulse sequences could be employed to get more insight in charge and spin dynamics of electrons and holes in self assembled QDs. 
\begin{figure}[h]
	\centering
		\includegraphics[width=1\textwidth]{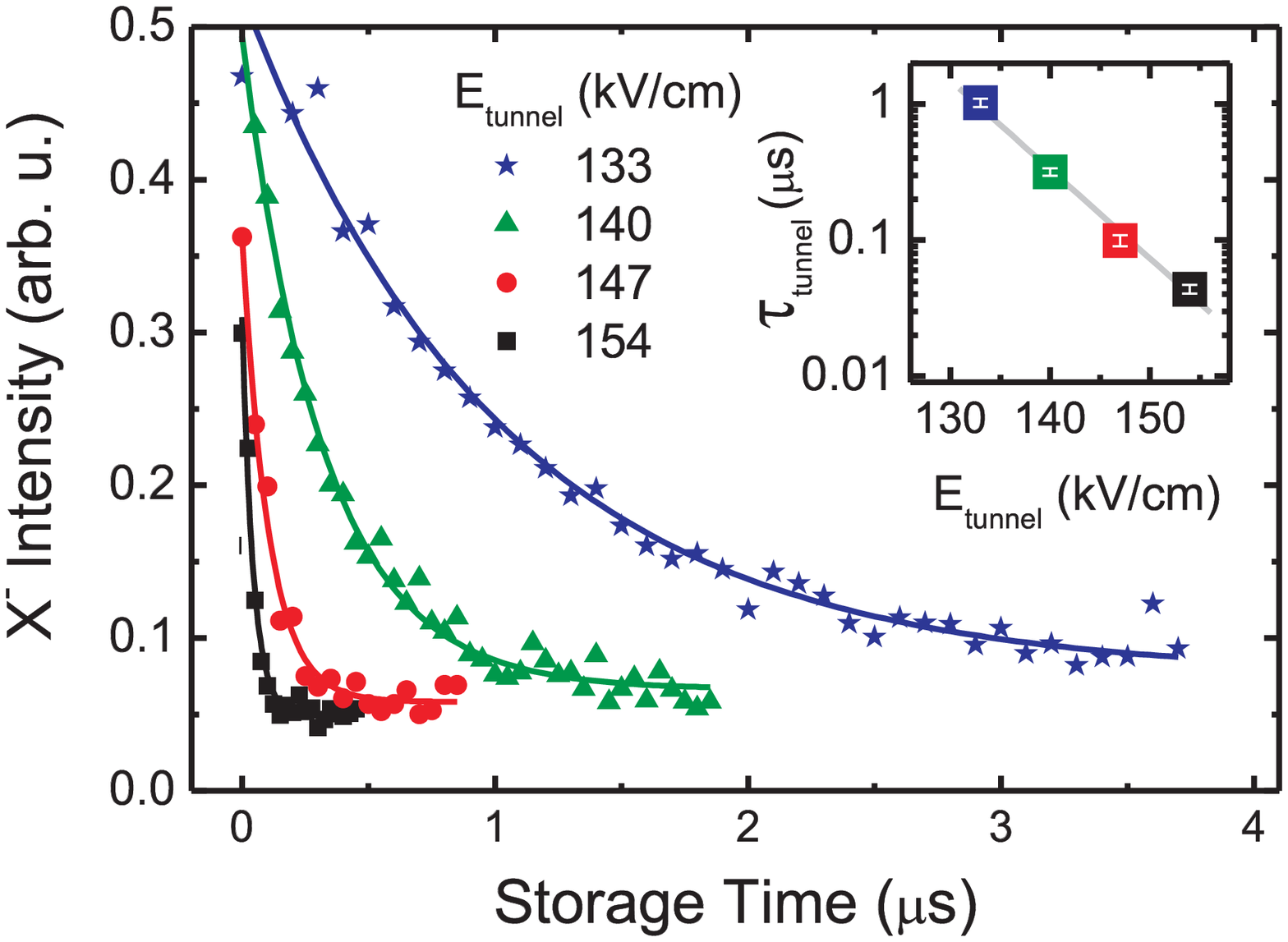}
	\caption{(color online) Time evolution of the X$^{-}$ intensity for several tunneling fields F$_{tunnel}$. Tunneling escape of the electron leads to an exponential decrease. The inset shows the extracted electron lifetimes. }
\end{figure}
\begin{acknowledgments}
The authors gratefully acknowledge financial support by the \textit{DFG} via \textit{SFB}-631 and the German Excellence Initiative via the \textit{Nanosystems Initiative Munich} (NIM).
\end{acknowledgments}
%

\newpage
\begin{thebibliography}\\

\bibitem{Hanson07} R. Hanson, L. P. Kouwenhoven, J. R. Petta, S. Tarucha, and L. M. K. Vandersypen, 
Rev. Mod. Phys. \textbf{79}, 1217 (2007).

\bibitem{Elzerman03} J. M. Elzerman, R. Hanson, J. S. Greidanus, L. H. Willems van Beveren, S. De Franceschi, L. M. K. Vandersypen, S. Tarucha, and L. P. Kouwenhoven, Phys. Rev. \textbf{B 67}, 161308 (2003).

\bibitem {Petta05} J.R. Petta, A.C. Johnson, J.M. Taylor, E.A. Laird, A. Yacoby, M.D. Lukin, C.M. Marcus, M.P. Hanson and A.C. Gossard, Science \textbf{309}, 2180 (2005).

\bibitem{Koppens06} F.H.L. Koppens, C. Buizert, K.J. Tielrooij, I.T. Vink, K.C. Nowack, T. Meunier, L.P. Kouwenhoven and M.K.  Vandersypen, Nature \textbf{442}, 766 (2006).

\bibitem{Kroutvar04} M. Kroutvar, Y. Ducommun, D. Heiss, M. Bichler, D. Schuh, G. Abstreiter and J.J. Finley, Nature , \textbf{432}, 81 (2004).

\bibitem{Heiss07} D. Heiss, S. Schaeck, H. Huebl, M. Bichler, G. Abstreiter, J. J. Finley, D. V. Bulaev and Daniel Loss, Phys. Rev. \textbf{B 76}, 241306(R) (2007).

\bibitem{Hoegele05} A. H\"ogele, M. Kroner, S. Seidl, K. Karrai, M. Atatüre, J. Dreiser, A. Imamolu, and R.J. 
arburton, Appl. Phys. Lett. \textbf{86}, 221905 (2005).

\bibitem{Ramsay08}  A. J. Ramsay, S. J. Boyle, R. S. Kolodka, J. B. B. Oliveira, J. Skiba-Szymanska, H. Y. Liu, M. Hopkinson, A. M. Fox, M. S. Skolnick,  cond-mat/07103738 (2008).

\bibitem{Atatuere07} M. Atat\"ure,J. Dreiser, A. Badolato, A. Imamoglu, Nature Physics, \textbf{3}, 2, 101 (2007).

\bibitem{Berezovsky06} J. Berezovsky, O. Gywat, F. Meier, D. Battaglia, X. Peng and D. D. Awschalom.  Nature Physics, \textbf{2}, 831 (2006).

\bibitem{Heiss08} D. Heiss, V. Jovanov, M. Bichler, G. Abstreiter, J. J. Finley, Phys. Rev. \textbf{B 77}, 235442 (2008).

\bibitem{Findeis01} F. Findeis, M. Baier, E. Beham, A. Zrenner, and G. Abstreiter, Appl. Phys. Lett. \textbf{78}, 2958 (2001). 
 
\bibitem{Fry00} P. W. Fry, J. J. Finley, L. R. Wilson, A. Lemaiˆtre, D. J. Mowbray, and M. S. Skolnick, M. Hopkinson, G. Hill, and J. C. Clark, Appl. Phys. Lett. \textbf{77}, (4344) (2000).
 
\end{thebibliography}
%
\newpage

Figure Captions\\
\\
FIG. 1: (color online) Schematic representation of the pulse sequence used in the storage experiments. The phases of the experiment are: (a) reset, (b) selective optical charging, (c) electron storage, (d) charge readout via time gated PL. \\
\\
FIG. 2: (color online) Photoluminescence - excitation measurements of a (a) charge neutral and (b) a charged QD. (c) Charging of the QD, observed by an increase of the X$^{-}$ intensity, can be achieved through discrete p-shell states of the QD.\\
\\
FIG. 3: (color online) Time evolution of the X$^{-}$ intensity for several tunneling fields F$_{tunnel}$. Tunneling escape of the electron leads to an exponential decrease. The inset shows the extracted electron lifetimes.\\
\end{document}